\documentclass[shortnote,twocolumn]{jpsj3}
\usepackage{txfonts}
\usepackage[dvipdfmx]{graphicx}
\usepackage{color}
\usepackage{dcolumn}
\usepackage{bm}
\usepackage{amsmath,amssymb}
\usepackage{ulem}

\begin{document}

\title{Magnetic excitations of Sr$_3$Ir$_2$O$_7$ observed by inelastic neutron scattering measurement}

\author{Masaki Fujita$^{1}$\thanks{fujita@imr.tohoku.ac.jp}, Kazuhiko Ikeuchi$^2$, Ryoichi Kajimoto$^3$, and Mitsutaka Nakamura$^3$ 
}
\inst{
$^1$Institute for Materials Research, Tohoku University, Katahira, Sendai 980-8577, Japan\\
$^2$Research Center for Neutron Science and Technology, Comprehensive Research Organization for Science and Society, Tokai, Ibaraki 319-1106, Japan\\
$^3$Materials and Life Science Division, J-PARC Center, Japan Atomic Energy Agency,  Tokai, Ibaraki 319-1195, Japan\\
} 

\abst{
In this short note, we report on the first inelastic neutron scattering (INS) study on high-energy magnetic excitations in Sr$_3$Ir$_2$O$_7$. We observed excitations between $\sim$80 meV and $\sim$180 meV. The peak position of the excitations is consistent with the dispersion of a single magnon determined by resonant inelastic X-ray scattering (RIXS) measurement. Thus, our results demonstrate the feasibility of INS for iridates, which has a large neutron absorption cross-section. 
}

\maketitle

Iridate oxides, Sr$_2$IrO$_4$ and Sr$_3$Ir$_2$O$_7$, are Mott insulators with spin-orbit entangled magnetic state bearing the effective total angular moment, $J_{\rm eff}=1/2$ \cite{Kim2008, Kim2009, Kim2012}. A new root for high-transition-temperature (high-$T_{\rm c}$) superconductivity is expected for iridate oxides~\cite{Wang2011, Watanabe2013}. Accordingly, their magnetic excitations has received significant attention analogous to the spin dynamics of high-$T_{\rm c}$ cuprate oxides. Resonant inelastic X-ray scattering (RIXS) is a powerful tool used to investigate magnetic excitations owing to its accessibility to high-energy regions even for small samples. The excitations of Sr$_2$IrO$_4$ and Sr$_3$Ir$_2$O$_7$ comprise a single magnon branch below 300 meV and a novel high magnetic mode (spin-orbit exciton with $J_{\rm eff}$ = 3/2) approximately 600 meV \cite{Kim2008, Kim2009, Kim2012}. 

Inelastic neutron scattering (INS) that directly measures the dynamical structure factor, S({\bf Q}, $\hbar\omega$), in the momentum ({\bf Q}) and energy ($\hbar\omega$) spaces is also an indispensable method for investigating spin dynamics. For cuprate oxides, INS provides a considerable amount of valuable information on magnetic excitations such as hourglass-shaped excitations\cite{Arai1999, Tranquada2004} and resonance excitations\cite{Rossat-Mignod1991, Mook1993} in the superconducting phase. 
In addition, ${\rm f}({\bf Q})$ determined after the evaluation of S({\bf Q}, $\hbar\omega$) can yield information on orbital character\cite{Ito1976} and spin-orbit entangled excitations in iridate oxides. 
Herein, ${\rm f}({\bf Q})$ is the magnetic form factor that is a Fourier transform of spin density distribution in real space. However, INS measurement of iridium oxides is challenging owing to the significant neutron absorption cross-section of the iridium nuclei. Report of excitations exceeding 100 meV in iridate oxides by INS is scarce following our review of literature, whereas observation below $\sim$40 meV was recently achieved for monolayer Sr$_2$IrO$_4$~\cite{Calder2018}. In this short note, we present an INS investigation of bilayer Sr$_3$Ir$_2$O$_7$. By using high-energy neutrons with a reduced absorption effect, we observed a magnon branch existing between $\sim$80 meV and $\sim$180 meV, and verified the consistency in dispersion with that obtained by RIXS.

For the experiment, we prepared single crystals by the flux-method. The maximum size of the crystals was 2 mm $\times$ 2 mm $\times$ 1.5 mm. Approximately $500$ crystals with a total mass of $2.1$ grams were assembled on aluminum plates (with 0.1 mm thickness) such that the crystallographic $c$-axis is perpendicular to each plate. We performed INS measurements on $4$ SEASONS at the Materials and Life Science Experimental Facility (MLF) at J-PARC, Japan. A 250 meV incident neutron energy ($E_{\rm i}$) was selected to determine the magnon excitations below 200 meV. 
This $E_{\rm i}$ with a chopper frequency of 250 Hz provided an energy resolution ($\Delta$$E$) of 12.6 meV at  $\hbar\omega$ = 150 meV, which is better than $\Delta$$E$ of $\sim$30 meV in RIXS measurements \cite{Kim2008, Kim2009, Kim2012}. 
The signals were collected for two days under 300 kW beam power operation at 6 K, that is significantly below the magnetic ordering temperature ($\sim$280 K)\cite{Cao2002}. To minimize the absorption effect, the crystals were placed with the $c$-axis parallel to the incident neutron beam. In this study, we use the tetragonal notation to describe the Miller index. 

	\begin{figure}[b]
	\begin{center}
	\includegraphics[width=52mm]{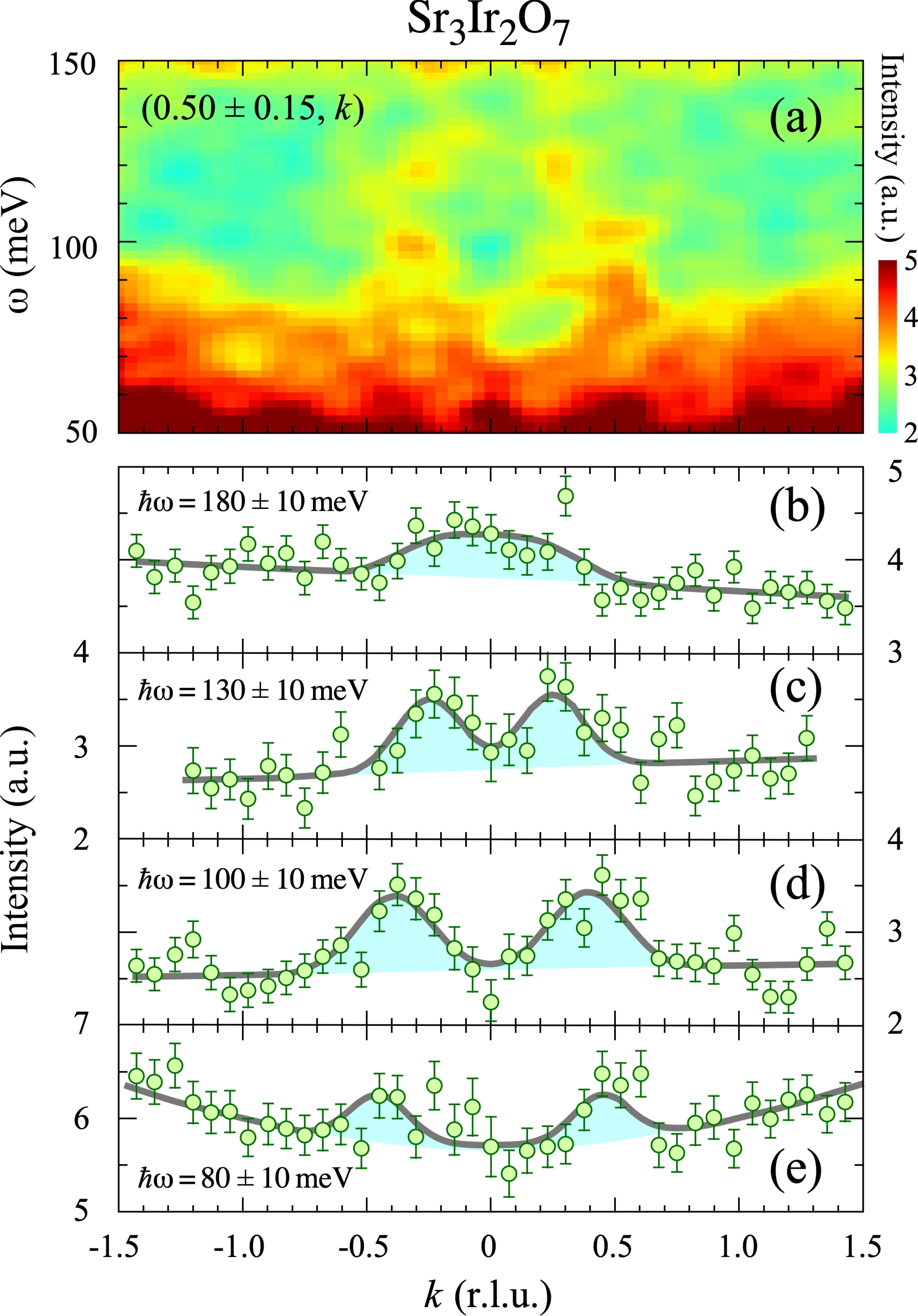}
	\caption{(Color online)~ (a) INS spectra of Sr$_3$Ir$_2$O$_7$ shown along the $k$-direction through (0.5, $\pm$ 0.5). The constant-energy spectra slices at (b) 180 $\pm$ 10 meV, (c) 130 $\pm$ 10 meV, (d) 100 $\pm$ 10 meV, and (e) 80 $\pm$ 10 meV. The gray lines are the fitted results by two equivalent Gaussian functions centered at (0.5, $\pm$ $k$) considering the linear or parabolic background. }
	\label{Const-E}
	\end{center}
	\end{figure}

Figure \ref{Const-E} shows the spectra along the $k$-direction through (0.5, $\pm$ 0.5) corresponding to the antiferromagnetic (AF) zone center (ZC). An arch-shaped signal centered at $k = 0$ is observed in Fig. \ref{Const-E} (a). The spectra sliced at $\hbar\omega = 180$ meV with an energy band of $\pm$ 10 meV (Fig. \ref{Const-E} (b)) shows a broad peak at $k = 0$. In decreasing $\hbar\omega$, the signal splits into two peaks that are symmetric around $k = 0$, and the distance between the peaks increases. Thus, the dispersive branch was successfully detected by neutron scattering measurements with sufficient beam flux.  
The peak positions at $\hbar\omega$ = 80 meV are close to ZC (Fig. \ref{Const-E} (e)), and no visible magnetic signal below 70 meV was observed, indicating the existence of a large energy gap in the spin excitations. 

From the figure, the scattering intensity for $|k| > 0.5$ is weak or absent, leading to the asymmetric spectra with respect to ZC ($k = \pm$ 0.5). Although the cause of the intensity unbalance is unknown, ${\rm f}({\bf Q})$ can be damped rapidly against {\bf Q} compared to that for cuprate oxides, owing to the extended nature of the $5d$ orbitals. In fact, in the neighborhood of ($1.5, 1.5$), which is ZC with a large $|{\bf Q}|$ value, the magnon branch was not observed. Thus, the intensity unbalance could be due to a particular ${\rm f}({\bf Q})$, and determining the intensity distribution via further INS with the quantitative analysis  can yield information on the orbital character.

\begin{figure}[t]
	\begin{center}
	\includegraphics[width=48mm]{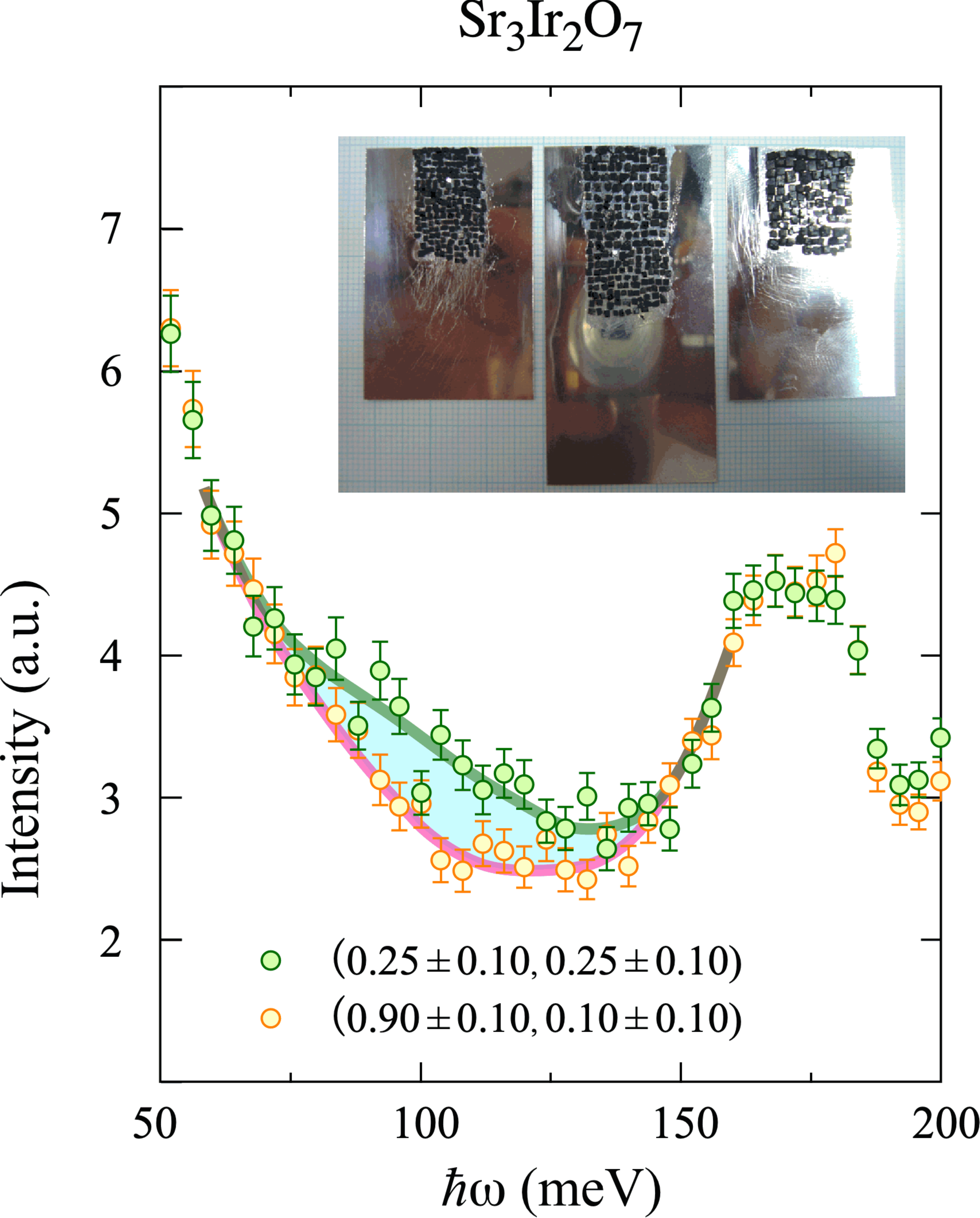}
	\caption{(Color online)~INS spectra slices at ($0.25, 0.25$) and ($0.9, 0.1$) with the $h$- and $k$-width of $0.1$. The inset is a photo of the assembled samples. }
	\label{Const-Q}
	\end{center}
	\end{figure}

To determine the gap energy at the AF zone boundary (ZB), we sliced the spectra at ($0.25, 0.25$) along the energy direction. (See Fig. \ref{Const-Q}.) For the comparison INS spectra at ($0.90, 0.10$), wherein the magnetic structure factor is weak, is shown as the background. The intensity around $100$ meV is higher at ($0.25, 0.25$) than that at ($0.90, 0.10$). From the difference in the intensities at the two positions, the peak-energy in the spectra at ($0.25, 0.25$) was evaluated to be $107(8)$ meV. This energy is consistent with the results reported in RIXS\cite{Kim2012}. Note that the peaks at $\sim$170 meV are independent of {\bf Q} and were observed even in the measurement without the crystals. Thus, this signal is background from outside the sample.

\begin{figure}[t]
	\begin{center}
	\includegraphics[width=75mm]{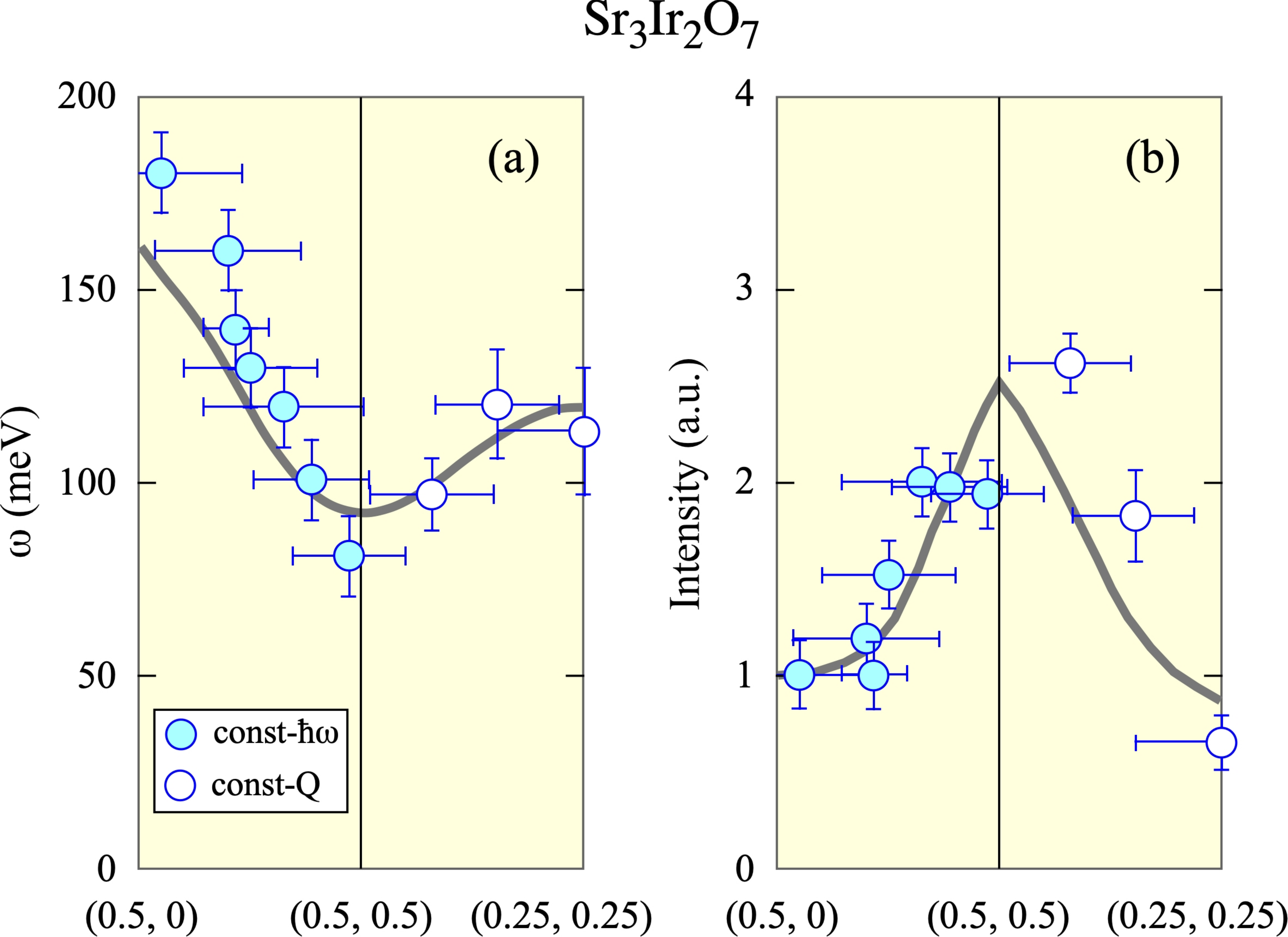}
	\caption{(Color online)~Momentum-dependence of (a) peak-position and (b) intensity in Sr$_3$Ir$_2$O$_7$. The gray lines are the results from RIXS [\ref{Kim2012}]. The intensity is not corrected with $|{\rm f}({\bf Q})|^2$ and the absorption coefficient.}
	\label{dispersion}
	\end{center}
	\end{figure}

In Fig. \ref{dispersion}, the momentum dependence of the peak-position and intensity are presented. The open (closed) circles represent the results obtained from the constant-momentum (constant-energy) spectra. The horizontal (vertical) bars for the open (closed) circles are the sliced momentum (energy) width of the spectra, and the vertical (horizontal) bars represent the evaluated peak-width in energy (momentum) in full-width at half-maximum. The peak positions follow a gray line representing the dispersion relation determined by RIXS. 
Figure \ref{dispersion} (b) shows the bare INS intensity that is not corrected with $|{\rm f}({\bf Q})|^2$ and the absorption coefficient depending on $\hbar\omega$. Accordingly, the values have systematic ambiguities against the Q position. However, the overall trend including the intensity maximum (minimum) around ZC (ZB) can be compared with the results of RIXS. 

From these results, we conclude the feasibility of the INS measurements on excitations in iridate oxides and that INS and RIXS detected identical excitations in the iridate oxides, as in the case of La$_2$CuO$_4$~\cite{Braicovich2010}. 
Although we did not measure the spin-orbit exciton found at $\sim$600 meV, the present results inspire a future investigation of this excitation by neu- tron scattering. INS measurement with polarization analysis can extract information such as the fluctuating component of spins, interference of spin and lattice dynamics and the orbital characters. The present experiment is foundational for such studies on iridium oxides, especially carrier doped compounds. 

In summary, we performed INS measurement on Sr$_3$Ir$_2$O$_7$, and successfully, observed the excitations between $\sim$80 meV and $\sim$180 meV. The peak positions agree well with the dispersion of a single magnon observed using RIXS. Therefore, the high energy neutrons aids in the investigation of the magnetic excitations of iridate oxides in the region of a hundred meV. 

The authors acknowledge K. Ishii and K. Tomiyasu for their useful remarks. The experiments at the Materials and Life Science Experimental Facility at J-PARC were performed under a user program (Proposal No. 2013A0052). M.F. is supported by a Grant-in-Aid for Scientific Research (A) (16H02125).

\end{document}